\documentclass[12pt,twoside,reqno,centertags]{amsart} 

\usepackage[tmargin=1in,bmargin=1in,lmargin=1in,rmargin=1in]{geometry}

\usepackage[utf8]{inputenc}
\usepackage[english]{babel}
\usepackage{subfigure} 
\usepackage{graphicx,hyperref,color}
\usepackage{rotating}
\usepackage[round]{natbib} 
\usepackage{booktabs} 
\usepackage{multirow} 
\usepackage{comment} 
\usepackage{subcaption}
\usepackage{caption}
\usepackage{amssymb}
\usepackage{amsmath} 
\usepackage{amsaddr}
\usepackage{epstopdf}
\usepackage{mathtools}
\usepackage{booktabs}
\usepackage{siunitx} 
\usepackage{dsfont}
\usepackage{microtype}

\newcommand{\E}{\mathds{E}} 
\newcommand{\re}{\mathds{R}}

\allowdisplaybreaks[4]

\newcommand\TT{\rule{0pt}{2.6ex}}       
\newcommand\BB{\rule[-1.2ex]{0pt}{0pt}} 

\usepackage{tgtermes}

\begin{document}

\title[Atheism and Cognitive Skills: Beta Regression]{Atheism and Cross-Country National Cognitive Skills: Evidence from a Beta Regression Analysis}
	
\author{Tatiene C.~ Souza}
	
\address{Departamento de Estat\'{\i}stica, 
		Universidade Federal da Para\'{\i}ba\\
		Jo\~{a}o Pessoa/PB, 58051-900, Brazil\\ 
        ORCID: 0000-0002-8873-0889\\
        E-mail: tatiene.souza@academico.ufpb.br} 
	
\author{Francisco Cribari-Neto}
	
\address{Departamento de Estat\'{\i}stica, 
		Universidade Federal de Pernambuco\\
		Recife/PE, 50670-901, Brazil\\ 
        ORCID: 0000-0002-5909-6698\\
		E-mail: francisco.cribari@ufpe.br\\
		(Corresponding author)} 

\begin{abstract}
This study performs a beta regression analysis to investigate how country-level cognitive skills---measured using harmonized large-scale international educational assessments and expressed on an IQ-type scale---are associated with the prevalence of atheism in 131 countries. We find that this association is nonlinear: at low levels of cognitive skills, increases are associated with lower expected levels of atheism, whereas beyond a threshold, they are associated with higher expected atheism rates. This turning point is estimated to occur at approximately 65 on the IQ-type scale. The strength and location of the relationship vary across regions. We estimate elasticities of expected atheism rates with respect to cognitive skills for all countries in the sample and introduce the concept of the atheism gap to identify countries where observed prevalence is substantially higher or lower than predicted by their conditioning variables.\\
\\
\textsc{Keywords}: Beta distribution, doubly bounded dependent variable, elasticity, intelligence, religion, religious belief 
\end{abstract}

\maketitle

\section{Introduction}\label{S:intro} 

Theories of secularization in sociology explain recent declines in religious belief as consequences of increased existential security and greater exposure to diverse worldviews---conditions under which religion becomes less central to identity and social organization \citep{Norris+Inglehart_2011}. Recent evidence suggests that these changes are driven largely by generational replacement, reflecting a gradual secular transition across societies \citep{Stolz+et-al_2025}. Secularization has advanced in many regions, particularly in Europe and North America, where younger cohorts exhibit lower levels of religious affiliation than older generations \citep{Inglehart_2021}.

In parallel, a growing body of research highlights the role of cognitive factors in shaping religious disbelief. Studies at the individual level generally report that individuals with higher cognitive ability are more likely to question and reject traditional religious beliefs \citep{Zuckerman+Li+Lin+Hall_2019, Kirkegaard+Dutton_2022}. However, this literature focuses primarily on individual-level measures, leaving open the role of aggregate cognitive characteristics in explaining cross-country variation in disbelief. In this study, we instead use country-level cognitive skills derived from international educational assessments, capturing average achievement-based performance and expressed on an IQ-type scale for comparability. While existing perspectives emphasize structural and generational mechanisms, they do not fully address the role of such aggregate cognitive factors in shaping cross-national patterns of religious disbelief.

Recent alternatives such as IQ.UK \citep{Gust+Hanushek+Woessmann_2024} and HLO.IQ \citep{Angrist+Djankov+Goldberg+Patrinos_2021} provide harmonized cross-national measures of cognitive skills. The measure used in this study does not correspond to psychometric IQ at the individual level, but is instead based on internationally standardized student achievement data harmonized onto a common scale and expressed in IQ units for comparability. Thus, it reflects country-level cognitive skills derived from educational performance rather than direct measures of intelligence. At the same time, such aggregate measures are closely related to broader socioeconomic and institutional conditions, which are explicitly controlled for in the empirical analysis. Building on this literature, the present study extends previous research by employing beta regression for bounded proportions, incorporating nonlinear and regional effects, and computing elasticity measures that quantify how responsive the expected prevalence of atheism is to changes in country-level cognitive skills.

Several cross-national studies report a negative association between cognitive ability and religiosity \citep{Rindermann_2018, Zuckerman+Li+Lin+Hall_2019}. Meta-analytic evidence suggests that this relationship operates partly through the mediating role of analytical reasoning. However, its strength is not uniform across social and cultural contexts, indicating that the association between country-level cognitive skills and religious disbelief may vary systematically across countries.

Lynn and Vanhanen (2012) provide one of the earliest cross-national analyses of the relationship between intelligence and religiosity, documenting a strong and nonlinear negative association between national IQ and religious belief. Their findings suggest that religiosity tends to decline more sharply once countries surpass an average IQ of approximately 85, although regional and cultural factors produce important deviations from this pattern.

\citet{Cribari+Souza_2013} investigated this relationship using data from 124 countries, identifying intelligence as a key predictor of atheism, alongside per capita income, trade openness, and Muslim-majority status. This study revisits and extends their analysis using updated data for 131 countries, adding seven cases to the original sample and incorporating more recent measures of both atheism and country-level cognitive skills. We employ beta regression to model the bounded share of religious disbelievers and introduce elasticity estimates that quantify how responsive the expected prevalence of atheism is to changes in cognitive skills, conditional on country characteristics. In doing so, we treat country-level cognitive skills as an important dimension of secularization alongside economic development and globalization.

Using a beta regression analysis, we examine whether the cognitive skills--atheism relationship is nonlinear and regionally heterogeneous across Latin America, Africa, and Muslim-majority countries. Our empirical results indicate that, although higher cognitive skills are generally associated with higher rates of atheism, the relationship may initially be negative in countries with very low cognitive skills. The peak responsiveness of atheism to cognitive skills also varies across regions, occurring later and more strongly in Muslim-majority countries. Finally, we estimate country-specific elasticities and identify countries whose observed atheism rates diverge substantially from model-based expectations, with Uruguay and Hong Kong displaying higher-than-expected atheism and Poland and Singapore lower-than-expected levels.

More specifically, this study contributes to the literature by: (i) updating the cross-country analysis with more recent data; (ii) introducing elasticity curves derived from a beta regression model fit to assess the nonlinearity and intensity of the association between cognitive skills and atheism; (iii) documenting regional heterogeneity in this relationship; (iv) using predictions from the beta regression model fit to identify countries whose atheism rates diverge substantially from model-based expectations; and (v) estimating the threshold beyond which the association between cognitive skills and atheism becomes positive, as well as the point at which this relationship begins to exhibit diminishing marginal returns.

The manuscript proceeds as follows: Sections~\ref{S:betaregression} and \ref{S:data} introduce the beta regression model and dataset, respectively. Section~\ref{S:empiricalanalysis} presents and discusses the findings. Section~\ref{S:conclusions} offers concluding remarks.

\section{The beta regression model}\label{S:betaregression}
	
Let $Y$ denote a random variable following the beta distribution with mean $\mu$ (i.e.\ $\E(Y) = \mu$) and precision parameter $\phi$. We write this as $Y \sim {\mathcal{B}}(\mu, \phi)$. The variance of $Y$ is $\mu(1-\mu)/(1+\phi)$, which decreases as $\phi$ increases. The beta density exhibits a variety of shapes depending on the parameter values. It can be symmetric, left-skewed, right-skewed, or assume a `J', `inverted J', or `U' shape.

Let $Y_1,\ldots,Y_n$ be independent random variables such that $Y_t \sim {\mathcal{B}}(\mu_t,\phi_t)$, $t=1,\ldots,n$. The beta regression model introduced by \citet{Ferrari+Cribari_2004} comprises two submodels, one for each parameter: 
\begin{equation*}
	g_1(\mu_t)=\sum_{j=1}^p x_{tj}\beta_j \eqcolon \eta_{1t} \,\, \text{ and } \,\,
	g_2(\phi_t) = \sum_{j=1}^q z_{tj}\gamma_j \eqcolon \eta_{2t},
\end{equation*}
where $\beta \coloneq (\beta_1,\ldots,\beta_p)^\top$ and $\gamma \coloneq (\gamma_1,\ldots,\gamma_q)^\top$ are vectors of regression coefficients ($p+q < n$), $\eta_{1t}$ and $\eta_{2t}$ are linear predictors, and $x_{t1},\ldots,x_{tp}$ and $z_{t1},\ldots,z_{tq}$ are observations of $p$ and $q$ explanatory variables that influence the mean responses and the precisions, respectively. Typically, $x_{t1} = z_{t1} = 1 \, \forall t$, so that the two submodels have intercepts. The link functions $g_1 : (0,1) \rightarrow \re$ and $g_2: \re_+ \rightarrow \re$ and their inverses must be strictly monotonic and twice continuously differentiable. Commonly used mean link functions are logit, probit, loglog, cloglog, and cauchit; in the precision submodel, the logarithm and square root links are regularly used \citep{Cribari+Zeileis_2010}.

Intuitively, the model separates two aspects of the response variable. The first submodel (for $\mu_t$) explains the expected value of the dependent variable. It uses a transformation of $\mu_t$, denoted $g_1(\mu_t)$, so that the left-hand side takes values on the real line, matching the scale of the linear predictor on the right-hand side. The second submodel (for $\phi_t$) explains how tightly the proportions cluster around their expected values \citep{Cribari+Zeileis_2010}. A higher $\phi_t$ indicates that countries with similar covariate values exhibit less response variability. Thus, while the mean submodel captures average tendencies, the precision submodel captures the degree of consistency across observations.
	
A distinctive feature of the beta regression model is its non-constant marginal effects. Unlike linear regression, the effect of a regressor on the mean of the dependent variable is not fixed. Instead, it depends on the regressor's coefficient, the values of all regressors, all regression coefficients in the mean submodel and the inverse mean link function. More formally, for $j=2, \ldots,p$,
\begin{equation*}
	\frac{\partial \mu_t}{\partial x_{tj}} = g_{1}^{-1^\prime}(\beta_1 + \beta_2 x_{t2} + \cdots + \beta_{p} x_{tp}) \, \beta_{j},
\end{equation*}
where $g_{1}^{-1^\prime}$ represents the derivative of the inverse mean link function with respect to $\eta_{1t}$. In the next section's empirical analysis, we will focus on the partial derivative of the mean prevalence of religious disbelief with respect to country-level cognitive skills (proxied by IQ-type measures).
	
The beta regression model's parameters are estimated using maximum likelihood, numerically maximizing the log-likelihood function with a Newton or quasi-Newton algorithm. For a more in-depth exploration of beta regression, please refer to \citet{Cribari+Zeileis_2010}. 
	
\subsection{The data}\label{S:data} 

This study models the proportion of atheists in 131 countries, using the following variables (sources and years in parentheses): NOGOD (2020, Pew Research Center): proportion of religious disbelievers, defined as the share of respondents who report not believing in God; IQ (Gust et al., 2024): country-level cognitive skills derived from harmonized international educational assessments, capturing average achievement-based cognitive performance and expressed on an IQ-type scale for comparability, rather than a direct measure of psychometric intelligence; INCOME (2022, World Bank): per capita Gross National Income (GNI) US dollars; OPEN (2021, World Bank): economic openness (sum of exports and imports as a percentage of GDP); MUSLIM (2024, World Population Review): percentage of Muslims; M: indicator for Muslim-majority countries; LA: indicator for Latin America; AF: indicator for Africa. While other variables were explored, the selected controls demonstrated the strongest explanatory power for cross-national variation in religious disbelief.

We adopt IQ.UK as an achievement-based proxy for country-level cognitive skills (referred to simply as IQ throughout the remainder of the paper). The variable IQ.UK corresponds to the national cognitive skill estimates developed by \citet{Gust+Hanushek+Woessmann_2024} and reported in the supplementary material of \citet{Warne_2022}. These estimates are constructed from harmonized international student achievement assessments, including PISA, TIMSS, and PIAAC, and are rescaled to the United Kingdom metric ($\text{mean}=100$). They are based on large-scale international educational assessments with nationally representative samples and standardized testing procedures, rather than on compilations of psychometric intelligence test samples. Although these assessments are not psychometric intelligence tests, they provide widely used cross-country indicators of national cognitive skills derived from educational achievement data. A substantial body of research shows that country-level results from international student assessments are strongly correlated with psychometric intelligence measures and capture closely related cognitive constructs at the population level (e.g., \citet{Rindermann_2018}; \citet{Henss_2024}). For ease of interpretation and comparability with the broader literature on intelligence and socioeconomic outcomes, we refer to this variable as IQ, while recognizing that it should be interpreted as an 
achievement-based proxy for country-level cognitive skills. 
It is important to emphasize that the IQ measure used in this study should not be interpreted as psychometric intelligence at the individual level, but rather as an aggregate index of cognitive performance derived from educational achievement data. As such, its values are expressed on an IQ-type scale for comparability, while reflecting relative performance in standardized international assessments rather than latent individual cognitive ability. This distinction is particularly relevant given ongoing criticisms of traditional ``national IQ'' datasets, as the measure used here is based on harmonized, large-scale international assessments with standardized methodologies, providing a more consistent and transparent basis for cross-country comparisons.

To account for economic development, we include GNI per capita, as wealthier nations tend to offer better education, healthcare, and scientific resources, which can reduce reliance on religious explanations \citep{Norris+Inglehart_2011}. GNI also captures levels of existential security, a central mechanism in sociological theories of secularization. Economic openness (OPEN) reflects a society's integration into global networks of trade and information. As globalization increases exposure to diverse worldviews, it may erode the dominance of traditional religious institutions and promote value pluralism \citep{Casanova_2001}.

Religious dominance is addressed through an indicator for Muslim-majority countries (M), where sociocultural pressures may limit religious skepticism \citep{Grim+Finke_2010}. In many such countries, the separation between religion and the state is less pronounced, and religious norms play a central role in shaping education, law, and public discourse. Moreover, the interpretation of sacred texts is often literal, and expressions of non-belief may face social or even legal restrictions. These structural and doctrinal features distinguish Muslim-majority societies from others and may condition how country-level cognitive skills relate to atheism prevalence.

To capture regional patterns, we also include Latin America (LA) and Africa (AF) indicators, as both regions have unique religious histories shaped by colonialism and local traditions. While Catholicism has historically dominated Latin America, Protestantism and secularism have gained ground \citep{Parker_2016}. In Africa, religious devotion remains strong, though urbanization and global influences are slowly shifting attitudes \citep{Warf_2014}. These regional indicators are not merely geographical; they represent distinct sociocultural regimes in which the relationship between country-level cognitive skills and atheism may be moderated by institutional, historical, or ideological factors. Incorporating these dummies enables the model to account for structural heterogeneity across global religious contexts.

The dataset employed in this study updates that of \citet{Cribari+Souza_2013} in two main respects. First, it expands the sample from 124 to 131 countries, incorporating additional cases for which reliable information has become available. Second, all key variables were replaced with more recent and methodologically improved indicators. 
Average country-level cognitive skills are now measured using the IQ.UK scores of \citet{Gust+Hanushek+Woessmann_2024}, which harmonize PISA, TIMSS, and PIAAC results into a cross-nationally comparable metric. The prevalence of atheism is drawn from the Pew Research Center’s 2020 global survey on belief in God, offering greater consistency across countries than earlier sources. These updates enhance coverage, comparability, and measurement reliability relative to the data used in the original study.

It is important to note that the concept of atheism employed here reflects the share of individuals reporting non-belief in God, as measured in international survey data. This measure captures what might be termed “soft atheism”---a lack of belief in a personal deity---rather than a full rejection of spirituality or metaphysical ideas. Consequently, cross-cultural differences in the meaning of “God” may affect interpretation. In non-theistic or polytheistic societies (e.g., East Asia or parts of Africa), the absence of belief in a monotheistic God does not necessarily imply irreligiosity. These cultural distinctions may contribute to regional outliers, particularly in East Asia, where non-theistic traditions such as Buddhism or Shinto coexist with high intelligence and comparatively low reported atheism. Therefore, the measure of atheism used in this study should be interpreted as an operational proxy for religious disbelief within a specific survey framework, rather than as a universally comparable indicator of irreligiosity across all cultural contexts. Nonetheless, it provides a consistent and widely used basis for cross-country comparisons in empirical research on religion.

The proportion of atheists (NOGOD) has a mean of 0.10 and a median of 0.04, ranging from 0.01 to 0.60, indicating that atheism prevalence varies considerably across countries. Table~\ref{tab:descriptive} presents descriptive statistics for continuous covariates. IQ has a mean of 84.15 and a median of 83.60, with values ranging from 55.00 to 108.60, indicating significant cross-country variation. The coefficients of variation reveal substantial relative heterogeneity across countries, particularly for INCOME and MUSLIM. The high coefficient of variation for INCOME is consistent with pronounced differences in economic development, whereas that for MUSLIM is consistent with the wide range of Muslim population shares, from countries with virtually no Muslim population to countries where Muslims constitute the vast majority of the population. OPEN also exhibits considerable dispersion, with its maximum value substantially exceeding the third quartile. These differences underscore the marked cross-country heterogeneity in the socioeconomic and demographic characteristics considered in the analysis.

\begin{table}[htb]
\caption{Descriptive statistics of the continuous explanatory variables.}
\label{tab:descriptive}
\begin{tabular}{lrrrr}
\toprule
& \multicolumn{1}{c}{IQ} & \multicolumn{1}{c}{INCOME} & \multicolumn{1}{c}{OPEN} & \multicolumn{1}{c}{MUSLIM} \\
\midrule
Minimum & 55.00 & 712 & 22.58 & 0.00\TT \\
First quartile & 75.40 & 5,336& 53.04 & 0.60\\
Median & 83.60 & 14,842 & 73.36 & 5.09\\
Mean & 84.15 & 22,664 & 86.26 & 23.75\\
Third quartile & 95.90 & 35,233 & 100.39 & 38.11\\
Maximum & 108.60 & 95,944 & 402.22 & 98.19\\
Coefficient of variation & 14.84 & 94.00 & 65.25 & 138.38\BB \\
\bottomrule
\end{tabular}
\end{table}

\section{Empirical results and discussion}\label{S:empiricalanalysis}

We employ beta regression models to analyze the proportion of atheists in 131 countries ($n=131$) as a function of country-level cognitive skills (proxied by IQ-type measures) and other relevant factors. Our primary objective is to determine whether a statistically significant association exists between country-level cognitive skills and the prevalence of religious disbelief. Additionally, we aim to characterize the functional form of this relationship. All estimations were conducted using the \textsl{betareg} package \citep{Cribari+Zeileis_2010} within the \textsf{R} statistical computing environment \citep{Manual_R}. The main computational effort involved estimating a large number of candidate models during model selection and assessing their adequacy using half-normal plots with simulated envelopes. For each plot, 100 artificial datasets were generated to construct the envelope. Despite this repeated simulation, the overall computational cost was modest.

\subsection{Model selection and estimation} 

Model selection was guided by the pseudo-$R^2$ measures proposed by \citet{Ferrari+Cribari_2004} and \citet{Nagelkerke_1991} (measures of overall model fit), the Akaike and Bayesian Information Criteria (AIC and BIC; balancing model fit with complexity), the RESET test of \citet{Ramsey_1969} (which tests the model's correct specification), half-normal residual plots with simulated envelopes (assessing model adequacy), and other residual plots. The envelopes, based on 100 simulations using the standardized weighted residual 2 proposed by \citet{Espinheira+Ferrari+Cribari_2008}, represent the 0.025 and 0.975 quantiles of the residual from the simulated datasets. We considered models with the percentage of Muslims in the population and models with a dummy variable indicating a majority-Muslim population. The best fits were obtained with the latter covariate. We considered indicator variables for different geographical regions. The best fit was obtained by considering regional effects associated with Latin America and Africa. After a careful process of selecting the variables, we have arrived at the following beta regression model:
\begin{align*}
\text{cloglog}(\mu_t) &= \beta_1 + \beta_2 \text{IQ}_t + \beta_3 \text{IQ}^2_t + \beta_4 \text{INT1}_t + \beta_5 \text{M}_t + \beta_6 \text{LA}_t+ \beta_7 \text{INT2}_t \\
\log(\phi_t) &= \gamma_1+ \gamma_2 \text{IQ}_t^{0.5}+ \gamma_3 \text{INCOME.K}_t,
\end{align*}
for $t = 1, \dots, 131$, where $\text{INCOME.K}_t$ is $\text{INCOME}_t/10^4$, $\text{INT1}_t = \text{INCOME.K}_t^2 \!\times\! \text{OPEN.L}_t^2$, and $\text{INT2}_t = \text{AF}_t \!\times\! \text{OPEN.L}_t^{1.5}$, with $\text{OPEN.L}_t = \log(\text{OPEN}_t)/10$. Here, $\text{cloglog}(\mu_t) = \log(-\log(1-\mu_t))$. Table~\ref{ta:estimate} presents the maximum likelihood estimates of the model parameters, along with the $p\text{-values}$ of the $z$-tests for individual regressor exclusion. We use $\hat{\beta}$ and $\hat{\gamma}$ to indicate the estimated coefficients in the mean and precision submodels, respectively.
	
\begin{table}[htb]
\caption{Parameter estimates and $p$-values of the individual $z$-tests for the fitted beta regression model (mean and precision submodels).}
\label{ta:estimate}
\begin{tabular}{lrrrr}
\toprule
& \multicolumn{1}{c}{$\hat{\beta}$} & \multicolumn{1}{c}{$p$-value} & \multicolumn{1}{c}{$\hat{\gamma}$} & \multicolumn{1}{c}{$p$-value} \\
\midrule
Intercept      &  $ 6.1975$  &  $0.0182$  &  $16.8228$  &  $<0.0001$ \TT \\  
$\text{IQ}$           &  $-0.3128$  &  $<0.0001$  &  &  \TT\\
$\text{IQ}^2$  &$0.0024$  &  $<0.0001$  &  &  \TT\\
$\text{INT1}$               &  $-0.0608$  & $<0.0001$   &  &  \TT\\
$\text{M}$               &  $-0.6230$ &  $0.0003$  &  & \TT \\
$\text{LA}$               &  $0.9710$ &  $<0.0001$  &  & \TT \\
$\text{INT2}$               &  $2.3079$  &  $0.0032$  &  &  \TT\\
$\text{IQ}^{0.5}$&     &        & $-1.5468$ & $<0.0001$  \TT  \\ 
$\text{INCOME.K}$&     &        & $0.2621$ & $0.0025$ \TT \BB \\ 
\bottomrule
\end{tabular}
\end{table}

All covariates are statistically significant at the 5\% significance level. The relationship between $\text{IQ}$ and the mean proportion of atheists is characterized by a quadratic association: in countries with lower average levels of cognitive skills (as proxied by $\text{IQ}$), an increase in $\text{IQ}$ generally is associated with a decrease in the expected proportion of atheists. However, this negative association weakens as cognitive skills rise, eventually reversing at higher $\text{IQ}$ levels. This will be further examined when we analyze $\partial \mu_t / \partial \text{IQ}_t$.  

The negative coefficient on the interaction between per capita income and trade openness ($\text{INT1}$) suggests that while both factors individually are associated with higher levels of secularization, their combined effect may be associated with a moderation of this pattern. In wealthier, open economies, religious institutions often play significant social roles, which may help sustain religious engagement despite the presence of factors typically associated with secularization. The variable $\text{M}$ (Muslim-majority countries) has a negative coefficient, indicating a lower expected proportion of atheists in these countries, whereas the positive coefficient for $\text{LA}$ (Latin America) is associated with a higher expected proportion of atheists. The positive coefficient for the interaction term $\text{INT2}$ suggests that within African countries, greater trade openness is associated with a higher expected proportion of atheists.  

In the precision submodel, the negative coefficient for $\text{IQ}^{0.5}$ indicates greater dispersion in the expected proportion of atheists across countries with higher levels of cognitive skills, while the positive coefficient for $\text{INCOME.K}$ suggests that higher income is associated with lower variability in these expected proportions. These opposing effects suggest a balancing pattern, shaping the distribution of atheism across nations. The ratio between the largest and smallest estimated precisions is approximately 62, confirming significant variation in precision.

The pseudo-$R^2$ measures proposed by \citet{Ferrari+Cribari_2004} and \citet{Nagelkerke_1991} are 0.69 and 0.70, respectively. Unlike the conventional $R^2$ in ordinary least squares regression, these measures should not be interpreted as the proportions of variation in the response explained by the model. Both measures take values in $[0,1]$, with larger values indicating a better fit according to the respective criteria. The pseudo-$R^2$ proposed by \citet{Ferrari+Cribari_2004} is defined as the squared correlation coefficient between $g_1(Y_t)$ and $\hat{\eta}_{1t}$, where the hat indicates evaluation at the maximum likelihood estimates. The pseudo-$R^2$ proposed by \citet{Nagelkerke_1991}, in turn, is based on a comparison between the maximized log-likelihoods of the fitted model and a reduced model containing intercepts only in both the mean and dispersion submodels. The model yields $\text{AIC} = -485.19$ and $\text{BIC} = -456.44$. These information criteria account for model fit and complexity and are primarily useful for comparing competing models fitted to the same data, with lower values indicating a preferable trade-off between fit and complexity.

The inclusion of the quadratic term in country-level cognitive skills (proxied by $\text{IQ}$) aims to capture the possibility that the relationship between cognitive skills and atheism is not strictly monotonic. In theoretical terms, low levels of cognitive skills may be associated with stronger adherence to traditional belief systems and limited exposure to alternative worldviews, which is consistent with a negative marginal association between cognitive skills and disbelief. As cognitive and educational resources increase, individuals may become more capable of abstract reasoning and less reliant on socially transmitted explanations of existence, which is consistent with a transition toward positive marginal associations. However, this relationship is unlikely to grow indefinitely. In high levels of cognitive skills or highly developed societies, atheism may reach a saturation point beyond which additional increases in cognitive skills or education remain associated with higher expected levels of disbelief, but at a progressively slower rate. This deceleration can be interpreted as reflecting the cultural normalization of secular worldviews, so that disbelief no longer signals cognitive differentiation but rather reflects broader processes of social conformity or ideological diffusion. Hence, the quadratic specification allows the model to capture a pattern in which the responsiveness of the expected proportion of atheists to cognitive skills increases up to a turning point and then gradually declines, consistent with the notion of diminishing marginal associations between cognitive skills and secularization.

\begin{figure}[htb]
\begin{minipage}[b]{0.38\textwidth} 
\includegraphics[width=\textwidth]{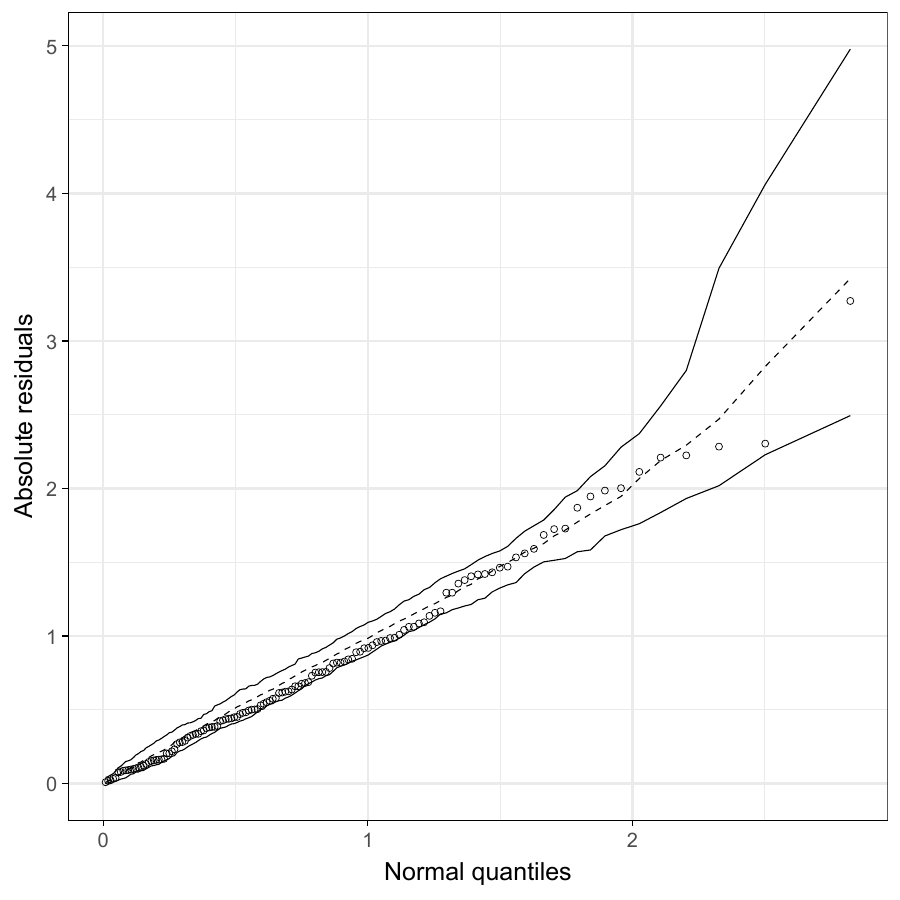}
\captionsetup{skip=0pt} 
\label{fig:env}  
\caption*{(a)}  
\end{minipage}
\begin{minipage}[b]{0.38\textwidth}
\includegraphics[width=\textwidth]{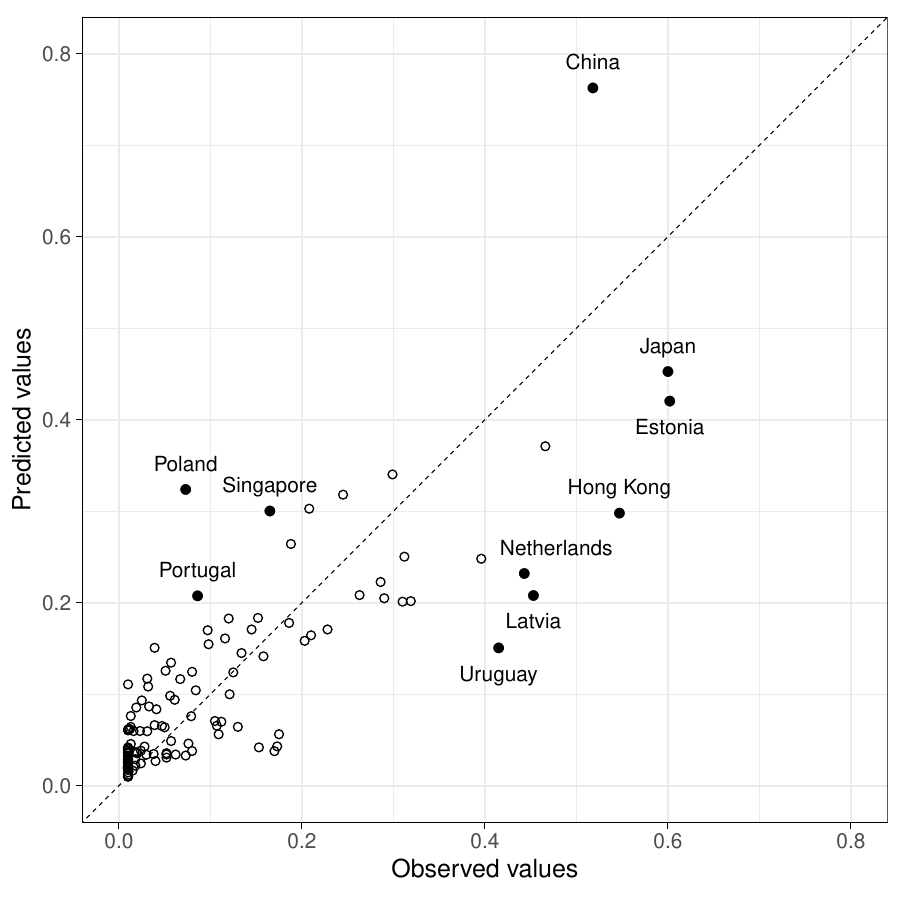}
\captionsetup{skip=0pt}  
\label{fig:predict_values}  
\caption*{(b)}  
\end{minipage}  
\caption{Residual normal probability plot with simulated envelopes (a) and observed versus predicted responses (b); the response variable is the proportion of atheists in 131 nations.}
\label{fig:env.predict_values}  
\end{figure}

The RESET specification test ($p\text{-value} = 0.398$) does not reject the null hypothesis of correct model specification; for details on the RESET test, see \cite{Pereira+Cribari_2014-2}. Figure~\ref{fig:env.predict_values} presents the residuals normal probability plot with simulated envelopes (left panel) and the observed versus predicted values plot ($y_t$ versus $\hat{\mu}_t = g_1^{-1}(\hat{\eta}_{1t})$, hats denoting evaluation at parameter estimates; right panel). All residuals fall within the simulated envelopes, confirming model validity. The right panel shows a strong alignment between predicted and actual atheist prevalence, though some countries deviate from expectations. These cases will be discussed later.    	
	
\citet{Rauber+Cribari+Bayer_2020} revisited the study of \citet{Cribari+Souza_2013} by estimating a beta regression model with a parametric link function for the mean. We also estimated our beta model using a parametric link for the mean with our dataset. However, the model with the cloglog link provided a better fit according to the AIC ($-485.19$ versus $-472.03$), BIC ($-456.44$ versus $-440.40$), and Nagelkerke's pseudo-$R^2$ ($0.70$ versus $0.67$). Therefore, the empirical analysis that follows is based on the model that employs the cloglog link in the mean submodel. 
	
\subsection{Elasticity curves}
	
As noted earlier, a primary goal of this paper is to examine the association between IQ-type cognitive skills and the prevalence of atheism. Specifically, we focus on how changes in IQ-type cognitive skills alter the mean proportion of atheists, holding all other relevant variables constant. To achieve this, we estimate responsiveness curves that show the expected change in the mean prevalence of atheists in response to variations in IQ-type cognitive skills. 

We evaluate five scenarios to assess how country-level cognitive skills (proxied by IQ-type measures) affect the mean proportion of atheists. Scenario~I represents non-Muslim countries outside Latin America and Africa ($\text{M}_t = 0$, $\text{LA}_t = 0$, $\text{AF}_t = 0$). Scenario~II focuses on Latin America ($\text{LA}_t = 1$). Scenario~III includes non-Muslim African countries ($\text{AF}_t = 1$). Scenario~IV examines Muslim countries outside Africa ($\text{M}_t = 1$), while Scenario~V considers Muslim nations in Africa ($\text{M}_t = 1$, $\text{AF}_t = 1$). In all cases, $\text{INCOME.K}_t$ and $\text{OPEN.L}_t$ are set to their median values, and $\text{IQ}_t$ is evaluated at specified values along its observed range.
		
Evaluating the independent variables' influence in relative terms is crucial. Elasticity, a fundamental economic concept, measures this sensitivity. The elasticity of the mean prevalence of atheists with respect to IQ-type cognitive skills is
\begin{equation*} 
\epsilon_{t} \coloneq \frac{\partial \mu_t}{\partial \text{IQ}_t} \!\times\! \frac{\text{IQ}_t}{\mu_t}, 
\end{equation*}  
which represents the percentage change in $\mu_{t}$ due to a 1\% increase in $\text{IQ}_t$.
For our model,
\begin{align*}
\frac{\partial \mu_{t}}{\partial \text{IQ}_{t}} = &\exp\left[\left( \beta_1 + \beta_2 \text{IQ}_t + \beta_3 \text{IQ}^2_t + \beta_4 \text{INT1}_t + \beta_5 \text{M}_t + \beta_6 \text{LA}_t + \beta_7 \text{INT2}_t\right) \right. \\
	& \left. - \exp\left( \beta_1 + \beta_2 \text{IQ}_t + \beta_3 \text{IQ}^2_t + \beta_4 \text{INT1}_t + \beta_5 \text{M}_t + \beta_6 \text{LA}_t + \beta_7 \text{INT2}_t \right) \right] \\
	& \times \left( \beta_2 + 2 \beta_3 \text{IQ}_t \right).
\end{align*}
All elasticities are computed replacing unknown parameters with their estimates.  

The elasticity of expected atheism with respect to IQ-type cognitive skills is computed based on the expected value of the dependent variable, rather than on the observed proportion of atheists. Specifically, for each country, the elasticity is evaluated at the fitted mean $\mu_t$, obtained from the estimated beta regression model.

Although each country contributes a single observation, the estimation of $\mu_t$ depends on the full sample through the model parameters. Therefore, the country-specific elasticities reflect how the expected level of atheism responds to changes in IQ-type cognitive skills, conditional on each country's characteristics and the information contained in the entire dataset.

\begin{figure}[htb]
	\includegraphics[width=0.65\linewidth]{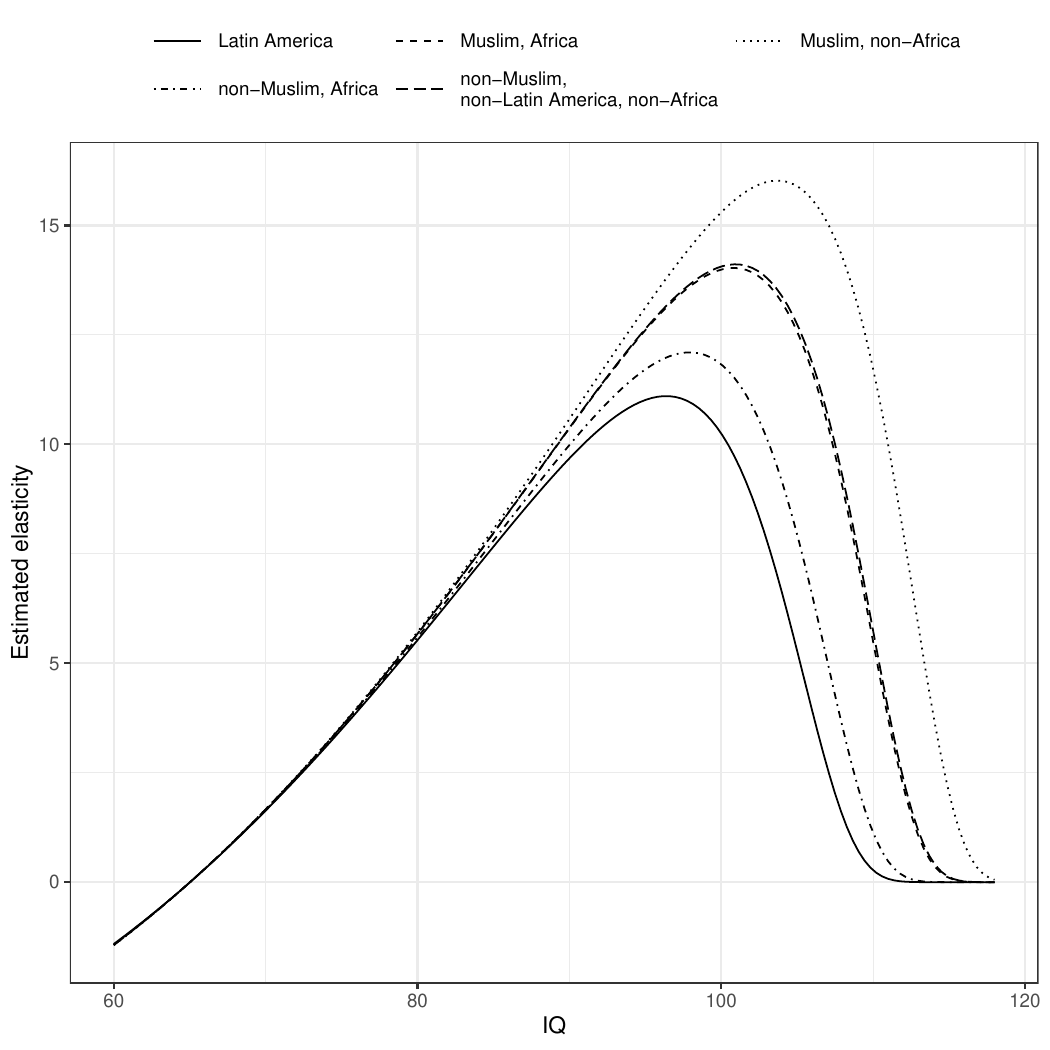}
	\caption{Estimated elasticities of the mean prevalence of atheists with respect to IQ-type cognitive skills across Scenarios~I to~V.}
	\label{fig:Elasticity}
\end{figure}

The elasticity curves presented below should therefore be interpreted as model-based quantities, reflecting changes in the expected prevalence of atheism rather than direct variations in observed country-level outcomes. Figure~\ref{fig:Elasticity} displays the estimated elasticity curves across the five scenarios. In Scenario~I (non-Muslim countries outside Latin America and Africa), elasticity peaks at 14.11 when IQ reaches 100.87, indicating that a 1\% increase in IQ is associated with an approximate 14\% increase in the expected prevalence of atheism. For Latin America (Scenario~II), the peak elasticity is 11.10 at IQ 96.20, indicating a lower level of responsiveness than in Scenario~I. Scenario~III (non-Muslim Africa) reaches 12.10 at IQ 97.76. Scenario~IV (Muslim-majority countries outside Africa) exhibits the highest peak elasticity (16.03) at IQ 103.60, while Scenario~V (Muslim-majority Africa) peaks at 14.03 at IQ 100.87, indicating a strong but slightly lower responsiveness than in Scenario~IV. Notably, elasticity is higher in non-African Muslim countries than in African ones, and the curves for Scenarios~I and V are strikingly similar.

These contrasts invite a sociocultural interpretation that helps contextualize the observed differences in elasticity patterns. Although both Latin America and the non-African Muslim world are characterized by strong religious institutions, their sociocultural roles differ. In Latin America, religion is often integrated into everyday social norms and civic life, creating a diffuse but stable cultural religiosity. Consequently, improvements in cognitive skills translate only gradually into secularization, leading to a lower peak elasticity. In contrast, religious institutions in many non-African Muslim countries exert centralized authority over doctrine and public behavior. When cognitive and informational pluralism expand---through education, media exposure, or global integration---individual responses may shift more abruptly, producing a higher and later elasticity peak. This contrast suggests that the observed nonlinearities reflect not merely institutional strength, but how different configurations of religious authority interact with processes of modernization and information diffusion. This interpretation is descriptive and contextual rather than causal, and the elasticity estimates should be understood as reflecting model-based, context-dependent responsiveness rather than a direct causal effect of IQ-type cognitive skills on secularization.

Because elasticity is defined with respect to a proportion rather than an absolute count, it can reach relatively high values in the intermediate range of the distribution of cognitive skills---where the predicted share of atheists is neither too low nor too high and the model’s sensitivity to IQ changes is greatest. To illustrate, consider a hypothetical country with an average IQ of 90, an atheism rate of 0.15, an elasticity of 10, and a population of one million inhabitants. A one-percent increase in mean IQ---from 90 to 90.9---is associated with an expected ten-percent increase in the atheism rate, raising it from 0.15 to 0.165. In absolute terms, the number of atheists would increase from 150{,}000 to 165{,}000---an increment of 15{,}000 individuals, corresponding to a rise of 1.5 percentage points in the population share. This illustrative example is intended solely to aid interpretation. It shows that, although an elasticity of 10 may appear large, its practical implications remain moderate: because the baseline proportion is relatively small, even a substantial relative change translates into a limited shift in the overall population composition.

Elasticity patterns are largely consistent up to IQ 85, diverging thereafter. Elasticity grows faster in cases where the peak is higher. The earlier this growth intensifies (starting at IQ 65), the more pronounced and delayed the subsequent decline.  

Negative elasticities appear in nations with very low levels of cognitive skills (measured on an IQ-type scale). One possible interpretation is that in impoverished settings, religion may provide psychological and social stability, with increases in cognitive skills reinforcing rather than challenging religious adherence. This contrasts with wealthier nations, where cognitive skills are often associated with secularization through exposure to diverse ideas. The nonlinear pattern may partly reflect the mediating role of formal education: in countries with very low average levels of cognitive skills (measured on an IQ-type scale), limited access to education---religious or secular---may restrict exposure to alternative worldviews, thereby reinforcing the social and psychological functions of religion. As education levels rise alongside cognitive skills, religious adherence may initially be strengthened by deeper engagement with doctrinal teachings, but beyond a certain threshold, cognitive and informational pluralism may tend to promote secularization. These interpretations should be viewed as plausible mechanisms rather than empirically identified causal pathways. It is important to note that this interpretation concerns predictive rather than causal relationships, since the empirical analysis identifies associations rather than direct causal mechanisms.

A key result is the identification of a threshold at which the elasticity becomes positive, indicating that increases in country-level cognitive skills (measured on an IQ-type scale) start to raise the expected prevalence of atheism. Fixing continuous regressors at their median values, we find that this tipping point occurs at approximately 65 on the IQ-type scale in all scenarios. Below this threshold, higher levels of country-level cognitive skills are associated with lower expected levels of atheism, providing suggestive evidence against the assumption that cognitive skills universally foster secularization. This pattern may reflect the influence of strong religious institutions and deeply rooted traditions in contexts with low levels of cognitive skills, which can shape belief formation and delay the emergence of secularizing effects associated with increases in cognitive skills.

The nonlinear pattern we uncover invites theoretical interpretation. \citet{Dutton2014-hm} argues that intelligence is generally negatively associated with religiosity because religion reflects evolved and instinctive forms of social cohesion, whereas higher cognitive ability facilitates the adoption of non-instinctive and secular worldviews. He further suggests that some political ideologies may operate as “replacement religions,” fulfilling similar psychological functions. Our finding of a nonlinear relationship between country-level cognitive skills and atheism is consistent with this perspective: as cognitive skills rise, the expected prevalence of atheism increases but eventually reaches a plateau, suggesting cultural or cognitive limits to secularization.

\subsection{Country-specific estimated elasticities}

\begin{figure}[p]
\centering
\includegraphics[width=1.0\linewidth]{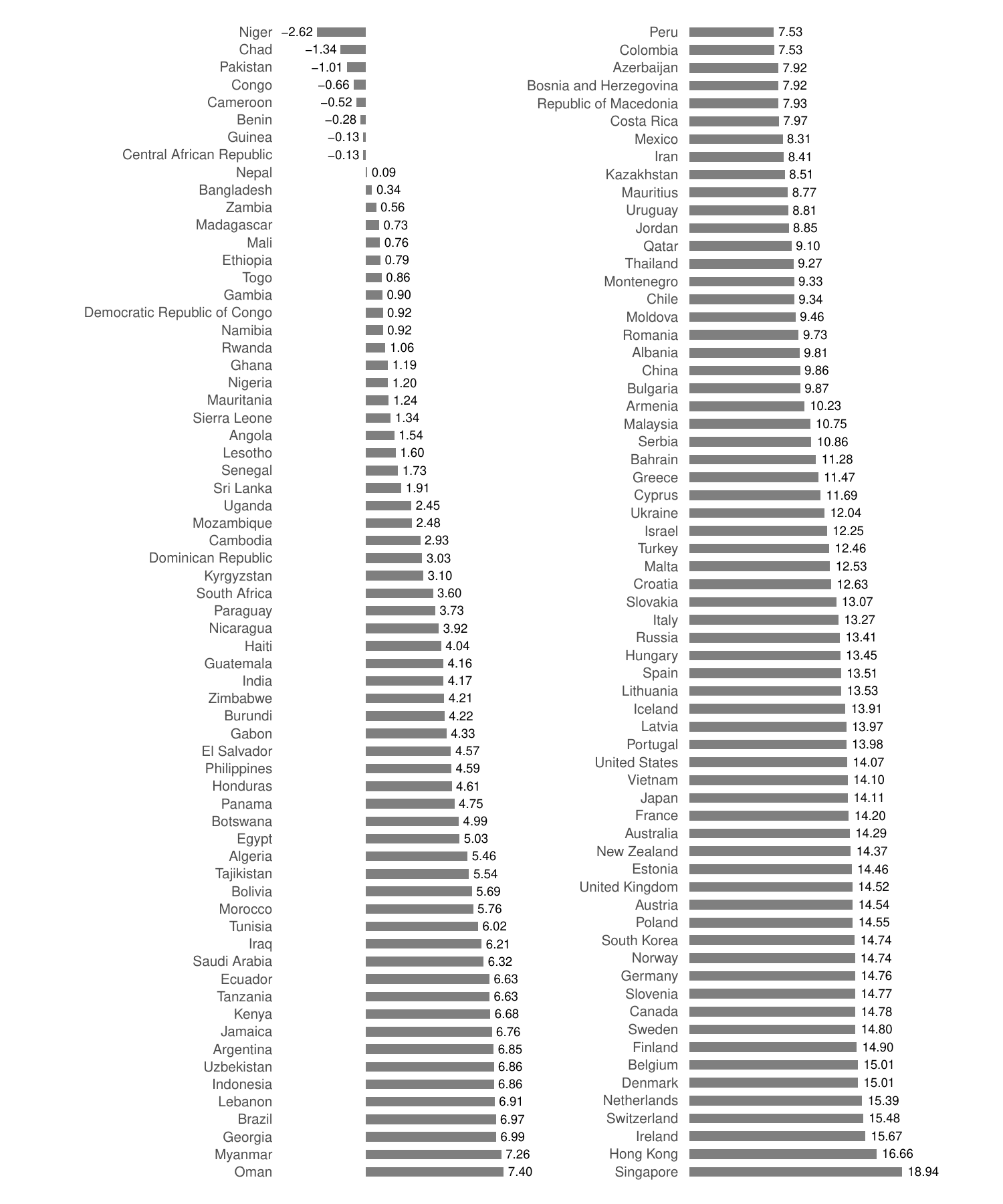}
\caption{Estimated elasticities for the 131 nations in the sample.}\label{F:barplot_elasticities_01}
\end{figure}

We estimated the elasticities of the mean prevalence of atheists with respect to country-level cognitive skills (proxied by IQ-type measures) using the values of the regressors from different countries and the parameter estimates. The average and median elasticities are 7.70 and 7.40, with a minimum of $-2.62$ and a maximum of 18.94. The first quartile is 3.67, and the third quartile is 12.85. The estimated elasticities are presented in ascending order in Figure~\ref{F:barplot_elasticities_01}.

The analysis of elasticities reveals substantial cross-country diversity among the 131 nations in the sample, with particular attention to the eight countries exhibiting negative elasticities. Seven of these are African countries, indicating a regional pattern in which increases in country-level cognitive skills (proxied by IQ-type measures) are associated with decreases in the prevalence of atheism. The negative elasticities range from $-2.62$ in Niger to $-0.13$ in Guinea and the Central African Republic. Pakistan is the only non-African country with negative elasticity. Among the 35 Asian countries in the sample, Pakistan stands out for having the lowest level of cognitive skills (as measured by IQ-type indicators), the lowest economic openness index (OPEN), and a per capita income (INCOME) below the first quartile of the Asian income distribution, suggesting that cultural and institutional religious factors may help sustain religiosity even as cognitive skills increase. 

Of the 131 countries analyzed, 37 are located in Africa. The mean elasticity for African countries is 2.22, whereas the median is 1.24, indicating that most African nations display relatively low elasticities, with a few higher values raising the mean. The highest elasticities in the region are observed in Mauritius (8.77), Kenya (6.68), Tanzania (6.63), and Tunisia (6.02). Even so, the lowest elasticities observed in Europe exceed the highest values recorded in Africa, with the exception of Mauritius. European countries with the lowest elasticities include Bosnia and Herzegovina (7.92), North Macedonia (7.93), Montenegro (9.33), and Moldova (9.46).

At the other extreme of the distribution are Singapore, Hong Kong, and several European countries with the largest elasticities, including Ireland (15.67), Switzerland (15.48), and the Netherlands (15.39). These results suggest that in such contexts even small increases in average country-level cognitive skills (measured on an IQ-type scale) can produce relatively large increases in the expected share of atheists, consistent with the idea that in highly developed societies religiosity tends to decline as cognitive and educational resources expand.

Elasticities in Latin America indicate a moderate relationship between country-level cognitive skills (measured on an IQ-type scale) and the prevalence of atheism. The mean and median values are 6.02 and 6.16, respectively. Chile (9.34), Uruguay (8.81), Mexico (8.31), and Costa Rica (7.97) display the highest elasticities, followed by Colombia and Peru (both 7.53). The lowest values occur in the Dominican Republic (3.03), Paraguay (3.73), and Nicaragua (3.92).

More broadly, elasticities tend to increase with socioeconomic development, although this relationship reflects a general tendency rather than a strict rule. In wealthier countries with greater access to education and resources, modest increases in average country-level cognitive skills (measured on an IQ-type scale) are often associated with comparatively larger changes in the expected prevalence of atheism. These patterns highlight the complex interaction between cognitive skills, socioeconomic development, and religious identity across different cultural contexts.

In addition to socioeconomic factors, cultural hostility toward atheism may influence how nonbelief is expressed and reported across societies. \citet{Nie_2023} shows that Christian nationalism significantly predicts anti-atheist prejudice even after controlling for religiosity and political ideology. In environments where atheism is socially stigmatized, individuals may be less willing to self-identify as nonbelievers, potentially contributing to underreporting.

Cultural and ideological dynamics may also explain cross-national differences in elasticity estimates. \citet{Webster+Duffy_2016} show that the negative association between intelligence and religiosity weakens in more recent cohorts. \citet{Dutton+VanderLinden_2017} attribute this pattern to the diffusion of atheism among individuals with lower cognitive skills. Likewise, \citet{WoodleyofMenie+Dunkel_2015} argue that ideological positions often signal conformity to prevailing cultural norms. Thus, the flatter intelligence--atheism elasticity observed in some highly developed or East Asian societies may reflect the normalization of secular worldviews, in which atheism becomes a socially accepted position rather than a marker of cognitive dissent.
	
\subsection{Countries with the largest deviations from expected atheism rates}

Our methodology produces estimates of the expected prevalence of atheism for each country ($\hat{\mu}_t$, $t=1,\ldots,131$), based on the conditioning variables included in the model. Comparing these predicted values with the corresponding observed rates ($y_t$) provides useful insights. To this end, we define the \textit{atheism gap} as 
\begin{equation*}
  \Delta_t \coloneq y_t - \hat{\mu}_t.
\end{equation*} 
Countries with positive (negative) values of $\Delta_t$ exhibit higher (lower) than expected atheism prevalence.

The ten countries with the largest positive gaps are Uruguay (0.2643), Hong Kong (0.2490), Latvia (0.2450), Netherlands (0.2110), Estonia (0.1816), New Zealand (0.1479), Japan (0.1473), Jamaica (0.1322), Mozambique (0.1299), and Botswana (0.1187). Three African countries appear in this group. The ten countries with the largest negative gaps are Poland ($-0.2508$), China ($-0.2447$), Singapore ($-0.1353$), Portugal ($-0.1216$), Iceland ($-0.1118$), Mauritius ($-0.1009$), Finland ($-0.0948$), Peru ($-0.0861$), Mexico ($-0.0776$), and Israel ($-0.0764$).

Explaining these deviations in detail falls beyond the scope of this study, but several illustrative cases highlight the role of historical, cultural, and institutional factors. Poland’s lower-than-predicted prevalence of atheism reflects the historically central role of Catholicism in national identity. The Church functioned as a unifying institution during periods of foreign domination and the communist era, reinforcing religious identification even as secularization advanced elsewhere in Europe. After communism, it also retained considerable political influence and remained closely intertwined with Polish nationalism \citep{Ramet_2017, Porter-Szucs_2011}.

China also displays a large negative gap despite its officially atheist ideology. Confucianism, Taoism, and folk religions coexist with secular worldviews and often include spiritual practices that fall outside strict definitions of atheism. Moreover, religion has remained resilient under communist rule. \citet{Yang_2011} describes a ``triple religious market'' in which regulated but persistent religious activity continues despite state restrictions. Similar dynamics were observed in the Soviet Union, where the campaign of ``scientific atheism'' failed to eliminate religious belief \citep{Froese_2004}.

Singapore also exhibits a lower-than-expected atheism rate despite very high country-level cognitive skills, income, and openness. Religious institutions remain influential through schools and community organizations \citep{Tan+Wu_2020}. At the same time, Singapore displays the highest elasticity of expected atheism with respect to country-level cognitive skills (18.94), suggesting that secularization could respond strongly to further increases in cognitive skills were these institutional constraints to weaken.

At the opposite extreme, Uruguay shows the largest positive atheism gap, reflecting its longstanding secular tradition and early separation of church and state \citep{DaCosta_2018}. Latvia and Estonia also display substantial positive gaps. In Estonia, secular and anti-clerical orientations became closely linked to national identity \citep{Gortfelder_2021}, while Latvia exhibits similar pragmatic and pluralistic attitudes toward religion \citep{Teraudkalns_2020}. Several African countries, including Botswana, South Africa, and Mozambique, also display positive gaps, possibly because religious pluralism, urbanization, and, in Mozambique, the legacy of civil conflict have weakened the dominance of traditional religious institutions.

More broadly, the Latin American context helps explain why the elasticity of atheism with respect to country-level cognitive skills tends to be moderate in the region. Although secularization is advancing, religious institutions continue to exert considerable social influence. \citet{Audette+etal_2020} show that churches in Latin America often cultivate civic skills and political participation through church-based activities, reinforcing the continued social relevance of religious organizations.

Our results also resonate with findings reported by \citet{Hirschle_2013}, who shows that economic growth is strongly associated with declining church attendance in Europe but not necessarily with declining religious belief. From a supply-side perspective, rising prosperity increases the availability of secular goods that substitute for some of the social functions of religion, reducing participation in religious institutions without immediately eroding belief.

Finally, countries with communist or post-communist histories present an informative contrast. Of the 131 countries analyzed, two are currently communist (China and Vietnam) and 24 are classified as post-communist. Both communist countries display negative atheism gaps ($-0.24$ for China and $-0.04$ for Vietnam). Among post-communist countries the average gap is also negative ($-0.016$), with particularly large negative values in Poland ($-0.25$) and Slovenia ($-0.08$). Nevertheless, there is substantial heterogeneity: Latvia and Estonia exhibit large positive gaps (0.25 and 0.18). This contrasts with the pattern reported by \citet{Dutton+VanderLinden_2017}, who argue—based on \citet{Lynn+Vanhanen_2012}—that former communist countries tend to be unusually atheistic relative to their average national IQs. Our results suggest that this historical pattern may have weakened or reversed, possibly reflecting post-communist religious revivals and generational changes.
	
\section{Concluding remarks}\label{S:conclusions}

This study performed a beta regression analysis on the relationship between average country-level cognitive skills (measured on an IQ-type scale) and the prevalence of atheism across 131 countries, reaffirming the positive, nonlinear correlation for nations with moderate to high levels of cognitive skills. Expanding on \citet{Cribari+Souza_2013}, we introduced a more nuanced regional segmentation, revealing significant variations in the intensity of the impact of cognitive skills on atheism across different cultural and religious contexts.

A key methodological contribution was the introduction of elasticity derived from the beta regression model fit as a refined metric to measure the sensitivity of expected atheism rates to country-level cognitive skills (measured on an IQ-type scale). Elasticity curves across five regional scenarios consistently showed an accelerating, peaking, and decelerating relationship. Notably, regions where the acceleration was more delayed exhibited higher peak effects, with Latin America reaching the lowest peak and Muslim countries outside Africa the highest.

A critical new finding was the identification of a threshold at approximately 65 on the IQ-type scale, below which increases in country-level cognitive skills are associated, within the model, with lower expected atheism prevalence. This insight challenges the assumption that cognitive skills universally foster secularization, suggesting that in contexts with low levels of cognitive skills, sociocultural factors may dominate belief formation.

We also computed elasticities from the beta regression fitted model for all sampled countries finding that eight countries, mostly in Africa, exhibit negative elasticities. The mean and median elasticities are around 7.5, implying that, within this range, a 1\% increase in country-level cognitive skills is associated with an approximate 7.5\% increase in the expected prevalence of atheism.  

We used the fitted model to identify countries with significant atheism gaps. Uruguay, Hong Kong, and Latvia show the largest positive gaps, while Poland and China exhibit the largest negative gaps.  

This study has some limitations. First, IQ-type measures of cognitive skills, while useful for cross-national comparisons, are subject to measurement error and cultural bias. Second, atheism was defined as disbelief in God, which may not fully capture broader secular orientations such as agnosticism or religious indifference. Third, the use of cross-sectional data prevents causal inference. Future research could employ longitudinal data to track how secular responsiveness to IQ-type cognitive skills evolves over time. Finally, despite accounting for potential nonlinearities, the elasticity approach remains based on aggregate-level data, which may obscure relevant individual-level variation in the cognitive skills--atheism relationship. Because the analysis relies on country-level aggregates, the results should not be interpreted as evidence about individual-level relationships between cognitive skills and atheism.

\section*{Disclosure statement}

No potential conflict of interest was reported by the authors. 

\section*{Funding}

This work was supported by Conselho Nacional de Desenvolvimento Cient\'{\i}fico e Tecnol\'{o}gico (CNPq) under Grant 304646/2023-7.

\section*{Data availability}

The country-level dataset used in the empirical analysis and an \textsf{R} script to reproduce the results are available at \url{https://github.com/cribari/atheism}. 

\section*{Authors' contributions} 

\textbf{Tatiene C.~Souza:} Conceptualization, Methodology, Software, Validation, Formal analysis, Investigation, Data Curation, Resources, Writing - Original Draft, Writing - Review \& Editing, Visualization. \textbf{Francisco Cribari-Neto:} Conceptualization, Methodology, Software, Validation, Formal analysis, Investigation, Resources, Writing - Original Draft, Writing - Review \& Editing, Visualization, Supervision, Project administration, Funding acquisition. 

	

\end{document}